\documentstyle[12pt,fleqn,epsfig]{article}
\begin{document}

\begin{center}

{\large \bf Gravitational lensing by wormholes}\\

\vspace{1cm}

 Tushar Kanti Dey\footnote[1]{tkdey54@rediffmail.com}   and Surajit
 Sen\footnote[2]{ssen55@yahoo.com}
 \\ Department of Physics \\ Guru Charan College\\
Silchar 788004, India \\

\end{center}

\vspace{.2cm}

\begin{abstract}
    We have investigated the gravitational lensing by two wormholes, viz.,
    Janis - Newman-Winnicour (JNW) wormhole and Ellis wormhole. The
    deflection angle in the strong field limit is calculated and
    various lens parameters of two wormholes are compared. It is
    shown that the JNW wormhole exhibits the relativistic
    images, while the Ellis wormhole does not have any relativistic
    images due to the absence of its photon sphere.
\end{abstract}

\vspace {2cm}


 \vfill \pagebreak

 \begin{center} \textbf{I. Introduction} \end{center}
\par
The gravitational lensing (GL) is regarded as one of the most
effective tool of probing a number of interesting phenomena of the
universe. Some of these phenomena charted recently are the existence
of various interesting astrophysical objects e.g.,  black holes,
super-dense neutron stars, exotic matter, wormholes, naked
singularity etc and the detection of these objects may eventually en
route to the possible connection between the quantum theory and
gravity. Since the classic observation of the bending of light in
1919, the general theory of relativity passes a number of stringent
tests mainly in the weak gravitational field limit. Out of these
tests, the observation of Einstein ring and the double or multiple
mirror images is the most profound example GL effect [1, 2, 3]. This
success leads to explore other extreme regime, namely, the GL effect
in the strong gravitational field limit, where the relativistic
images with a separation of the order of a few micro-arcseconds has
been predicted [4 - 7]. Out of various intriguing objects mentioned
above, recently the wormhole has renewed wide attention because of
its own right. Originally proposed by Morris and Thorne in 1988 [8],
the existence of this wormhole cannot be ruled out by the general
theory of relativity, although its feasibility requires the
existence of the exotic matter [8 - 10]. Therefore it is interesting
to study the GL effect caused by the wormhole particularly in the
strong gravitational field which is essentially related with other
important issues such as the evidence of the exotic matter,
formation of caustic etc [11, 12].
\par
The GL effect, either in the weak or strong gravitational field,
primarily involves the solution of the null geodesic equation. Its
solution in terms of the angle of deflection of light can be
expressed by the integral [13]

\begin{equation}
    \hat{\alpha} (r_0) =
2 \int_{r_0}^\infty \sqrt{A(r)/D(r)} \sqrt{\left[ \left(
\frac{r}{r_0}\right) ^2\frac{D(r)}{D(r_0)} \frac{B(r_0)}{B(r)}
 -1\right] ^{-1}} \quad \frac{dr} {r}  -  \pi,
\end{equation}
\noindent where $A(r)$, $B(r)$ and $D(r)$ be coefficient of the
space-space, time-time and angular parts of the generic line element
which reduce to unity in the asymptotic limit of r and $r_0$ be the
distance of closest approach of light to the compact object which
causes the lensing effect. It follows from the simple geometry that
the above integral in Eq.(1) is related to the basic lens parameters
as [4, 5]

\begin{equation}
tan{\beta}=tan{\theta}-\alpha(r_0),
\end{equation}
\noindent
where,

\begin{equation}
\alpha(r_0)=\frac{D_{ds}}{D_s}[\tan
\theta+\tan(\hat{\alpha}(r_0)-\theta)].
\end{equation}

\noindent Here $\beta$ and $\theta$ be the source and the image
position measured from the optic axis, $D_{ds}$ and $D_s$ be the
source-lens and source-observer distance respectively. The total
magnification of a circularly symmetric GL is given by [5]
\begin{equation}
\mu=\left(\frac{\sin{\beta}}{\sin{\theta}}
\frac{d\beta}{d\theta}\right)^{-1}
\end{equation}
and the tangential and radial magnification are given by

\begin{equation}
\mu_t=\left(\frac{\sin{\beta}}{\sin{\theta}}\right)^{-1} , \qquad
\mu_r=\left(\frac{d\beta}{d{\theta}}\right)^{-1}.
\end{equation}

\noindent Thus calculation of various lens parameters amounts to the
evaluation of the bending angle given by Eq.(1). In the recent past
several attempts have been made to calculate the elliptical integral
mainly by Virbhadra and Ellis [4, 5], Virbhadra, Narasimha and
Chitre [6], Bozza [7], Nandi et al [9], Amore and Diaz [14] and
others [15]. Out of these approaches recently Amore and Diaz (AD)
have proposed an elegant method to calculate the deflection angle
even in the strong gravitational field limit [14]. Their method
relies primarily on two approximations: firstly, they have used the
linear delta function technique to approximate the above integral
with a ansatz potential and then, secondly, they use the principle
of minimal sensitivity (PMS) to minimize the parametric dependence
to the deflection angle in the strong field limit. The primary
objective of the present work is to study the GL effect caused due
to the Janis-Newman-Winnicour (JNW) [5, 9, 10, 16, 17] and Ellis
wormhole [8, 18] in the strong gravitational field using the
methodology developed by AD.

\par
The remaining Sections of the paper are organized as follows: in
Section-II we review the essence of the approach of AD to develop
the general formalism of solving Eq.(1) in the strong gravitational
field. Section-III discusses the explicit calculation of the bending
angle in the strong and weak field limit for these wormholes. In
Section-IV we compare basic features of two wormholes and discuss
their magnifications. We conclude by summarizing the main results in
Section V and discuss the outlook.

\pagebreak

\begin{center} \textbf{II. Formalism} \end{center}

To find out the deflection angle following the procedure outlined by
Amore and Diaz [14], we change the variable $r=r_0/z$ and define the
potential function

\begin{equation}
V(z) = z^2 \frac{D(r_0/z)}{A(r_0/z)}-
 \frac{D^2(r_0/z) B(r_0)}{A(r_0/z) B(r_0/z) D(r_0)} +
 \frac{B(r_0)}{D(r_0)}.
\end{equation}

\noindent Using Eq.(6),  the  Eq (1) can be expressed as

\begin{equation}
\hat{\alpha}(r_0) = 2  \int_0^1 \frac{dz}{\sqrt{V(1) - V(z)}} - \pi.
\end{equation}
\noindent To evaluate the integral in Eq.(7) AD have used the linear
delta shifted potential as

\begin{equation}
V_{\delta}(z)=V_{0}(z)+\delta(V(z)-V_0(z)),
\end{equation}

\noindent where $V_0(z)=\lambda z^2$ be the ansatz potential. The
potential in Eq.(6) can be expressed as

\begin{equation}
V(z)= \sum\limits_{n = 1}^\infty  {v_n z^n },
\end{equation}

\noindent where $v_n$ be the coefficient of $z$ which are to be
determined. Plucking back Eq.(9) into Eq.(7) a straight forward
algebra gives the deflection angle to be

\begin{equation}
\hat{\alpha}^1(r_0)=\frac{3\pi}{2\sqrt{\lambda}}-\frac{1}{\lambda^{3/2}}\sum\limits_{n
= 1}^\infty  {v_n\Omega_n}-\pi,
\end{equation}
\noindent where the superscript $1$ corresponds to the first order
expansion of the integrand in Eq.(7) and $\Omega_n$ is given by

\begin{equation}
\Omega_n = \sqrt{\pi}\frac{\Gamma(n/2 +1/2)}{\Gamma(n/2)}.
\end{equation}
\noindent
 In Eq.(10), the unknown parameter $\lambda$ is obtained by the PMS,

\begin{equation}
 \left. \frac{\partial\hat{\alpha}^{1}}{\partial
\lambda}\right|_{\lambda=\lambda_{PMS}} =0,
\end{equation}
\noindent and the expression of the PMS approximated deflection
angle is given by
\begin{equation}
\hat{\alpha}_{PMS}^1(r_0) = \pi \left[{\sqrt{\frac{\pi }{2\rho}}-1}
\right],
\end{equation}

\noindent where

\begin{equation}
\rho=\sum\limits_{n=1}^\infty {v_n\Omega_n},
\end{equation}

\noindent be the transformed potential which needs to be calculated
for the metric in consideration. Eq.(13) is the form of the
deflection angle which was originally derived by AD and in next
Section we proceed to calculate it for two distinct wormhole
configurations.

\begin{center} \textbf{III. Calculation of deflection angle due to
wormholes} \end{center}

\textbf{ a) Janis-Newman-Winnicour wormhole}

\par
The matter with non-vanishing exotic scalar charge is described by
the Janis-Newman-Winnicour (JNW) metric

\begin{equation}
 ds^2  = \left( {1 - \frac{{2m}}{r}} \right)^\gamma dt^2  - \left(
{1 - \frac{{2m}}{r}} \right)^{ - \gamma } dr^2  - r^2 \left( {1 -
\frac{{2m}}{r}} \right)^{1 - \gamma } (d\theta ^2  + \sin ^2 \theta
d\phi ^2 ),
\end{equation}

\noindent where $\gamma = \frac{M}{m}$ with $M$ be the Arnowitt,
Deser and Misner (ADM) mass related to the asymptotic scalar charge
$q \left(= m \sqrt{\frac{2(1-\gamma^2)}{\kappa}}\right)$ by
$M^2=m^2-\frac{1}{2}\kappa q^2$. For the EMS (Einstein Massless
Scalar) theory, the solution of the covariant scalar solution
$\Phi(r)$ is related with the ADM mass by $\Phi(r) =
\sqrt{\frac{1-\gamma^2}{2\kappa}}\ell n[1-\frac{2m}{r}]$ where
$\kappa ( > 0 )$ represents the matter-scaler field coupling
constant. Several authors have calculated the deflection angle for
$\gamma\leq1$ with the scalar charge is real [6, 7], however for
$\gamma>1$, which corresponds to the JNW wormhole, the scalar charge
is a complex quantity [9, 10].

\par
The potential function $V(z)$ in Eq.(6) can be easily read off for
the JNW metric as

\begin{equation}
V(z) = z^2 \left( {1 - \frac{{2mz}}{{r_0 }}} \right) - \left( {1 -
\frac{{2mz}}{{r_0 }}} \right)^{2(1 - \gamma )} \left( {1 -
\frac{{2m}}{{r_0 }}} \right)^{2\gamma  - 1}  + \left( {1 -
\frac{{2m}}{{r_0 }}} \right)^{2\gamma  - 1}
\end{equation}

\noindent and it can be expanded around $z = 0$ to give

\begin{equation}
V(z) \approx v_1 z + v_2 z^2 + v_3 z^3 + O\left(z^4\right),
\end{equation}

\noindent where
\begin{equation}
 v_1  =  {\frac{{4m}}{{r_0 }}\left( {1 -
\frac{{2m}}{{r_0 }}} \right)^{  2\gamma -1}  (1 - \gamma )},
\end{equation}

\begin{equation}
 v_2  =  {1 - \frac{{4m^2 }}{{r_0^2 }}\left( {1 - \frac{{2m}}{{r_0
}}} \right)^{ 2\gamma -1 } (1 - \gamma )(1 - 2\gamma )},
\end{equation}

\begin{equation}
 v_3  =  - \frac{{2m}}{{r_0 }} - \frac{{16m^3 }}{{3r_0^3 }}\left(
{1 - \frac{{2m}}{{r_0 }}} \right)^{ 2\gamma -1 } \gamma (1 - \gamma
)(1 - 2\gamma ),
\end{equation}

 \noindent respectively. Using Eqs.(11) and (17), the transformation
 potential
 $\rho$ can be easily read off from Eq.(14)
and the PMS approximated deflection angle in Eq.(13) is given by
(for convenience, we have dropped the superscript $1$ and the
subscript PMS.)
\begin{equation}
\hat \alpha (r_0) = \pi \left[ {\sqrt {\frac{{3\pi (r_0  - b)r_0^2
}}{{3r_0 (r_0  - b)(\pi r_0  - 4b) + b(1 - b/r_0 )^{2\gamma } (1 -
\gamma )\{ 12r_0 ^2  - (1 - 2\gamma )(3 \pi b r_0   + 8 \gamma b^2
)}}} - 1} \right].
\end{equation}

\noindent For $\gamma = 1$, the JNW metric reduces to the
Schwarzschild metric and we obtain the deflection angle to be
$\hat{\alpha}(r_0) = \pi (\frac{1}{\sqrt{1-4 b/ \pi r_0}}-1)$, while
for $\gamma = 2$, we get the deflection angle corresponding to the
JNW wormhole

\begin{equation}
\hat {\alpha} (r_0) = \pi \left[ {\sqrt {\frac{{3\pi (r_0  - b)r_0^6
}}{{3br_0 ^5 (r_0  - b)(\pi r_0  - 4b) - 3b(r_0  - b)^4 (16b^2  +
3\pi br_0  + 4r_0^2 )}}}  - 1} \right].
\end{equation}
\noindent At large distance the deflection angle in the weak field
limit is given by

\[
 \left. {\hat \alpha }(r_0) \right|_{r_0  \to \infty }
 = \frac{{2b\gamma }}{{r_0 }} + \frac{{b^2 }}{{2\pi r_0 ^2 }}\left[ {12\gamma ^2
 - \pi (4 - \pi )(1 - 3\gamma  + 2\gamma ^2 )} \right] \]

 \[ + \frac{{b^3 }}{{6 \pi ^2 r_0 ^3 }}[ {3\pi ^3 (1 - \gamma )(1 - 2\gamma )^2
  + 120\gamma ^3  - 72\pi \gamma (1 - 3\gamma  + 2\gamma ^2 )}\]
\begin{equation}
   - 2\pi ^2 (6 - 37\gamma
  + 69\gamma ^2  - 38\gamma ^3 )]  + O\left( {\frac{1}{{r_0 }}}
\right)^{4} \\
 \end{equation}

\textbf{ b) Ellis wormhole }
\par
The Ellis wormhole is described by the metric [9, 13, 17]

\begin{equation}
ds^2 = dt^2 - dr^2 - (r^2 + a^2)(d\theta^2 + sin^2 \theta d\phi^2)
\end{equation}
\noindent and from Eq.(6) the potential function $V(z)$ is found to
be
\begin{equation}
V(z) = \left(\frac{r_0^2 - a^2}{r_0^2 + a^2}\right) z^2 +
\left(\frac{a^2}{r_0^2 + a^2}\right) z^4.
\end{equation}
\noindent It follows from Eq.(13) that the deflection angle is given
by
\begin{equation}
\hat{\alpha}(r_0) = \pi\left[ \sqrt{\frac{2(r_0^2 + a^2)}{2
r_0^2+a^2}} -1\right],
\end{equation}
\noindent and in the weak field limit it becomes

\begin{equation}
\hat{\alpha}(r_0)|_{r_0 \to \infty }  = \frac{{a^2 \pi }}{{4r_0^2 }}
- \frac{{5a^4 \pi }}{{32r_0^4 }} + \frac{{13a^6 \pi }}{{128r_0^6 }}
- \frac{{141a^8 \pi }}{{2048r_0^8 }} + O\left( {\frac{1}{{r_0 }}}
\right)^{10}.
\end{equation}
\noindent Eqs.(21) and (26) gives the expression of the deflection
angles in the strong field which is useful to calculate various lens
parameters for these wormholes.

\begin{center} \textbf{IV. Numerical results} \end{center}
\par

To explore the physical content, let us proceed to analyze the
deflection angle and various lens parameters for two wormholes
numerically. It is easy to see that for the JNW wormhole, the radius
of the photon sphere is given by $r_{ph}^{JNW}=4.27 m$ for
$\gamma=2$ which is less than the value predicted by the photon
sphere equation [5]. This result may be improved by including the
higher order calculation recently developed by Amore et al [20].
Fig.1 shows the variation of the deflection angle with the closest
distant of approach for the JNW wormhole. We note that the
deflection angle asymptotically goes to infinity when $x_0$
($=\frac{r_0}{2m}$) approaches to the photon sphere. To compare this
with the Ellis wormhole, Fig.2 depicts the plot of deflection angle
with closest distant of approach for this wormhole for different
values of the throat parameter $a$. It follows that, unlike the JNW
case, the deflection angle attains a critical maximum value of $\hat
\alpha (0) = 1.30$ radian ($74.6$ degrees). This manifestly shows
the non-existence of the photon sphere for the Ellis wormhole
indicating the absence of the relativistic images for this wormhole.
This result for the Ellis wormhole also follows from the photon
sphere equation [5] and it contradicts with the results of Perlick
[19] who argued for the existence of its relativistic images.
\par
To calculate the separation between the leading relativistic images
for a typical JNW metric we take $M=2.8 \times 10^6 M_ \odot$ and
the the lens-observer distance $ D_d = 8.5 {\rm{ kpc}}$.
Furthermore, we consider that the lens is situated at half way
between the source and the observer, i.e., $ D_{ds} /D_s  = 1/2$ and
the throat parameter to be $a=10$ km. For different values of
$\gamma$, Table-I illustrates closest approach distance $x_0$,
Einstein ring ($\theta_E$) with corresponding deflection angle
($\alpha_E$) and the separation between the first relativistic image
($\theta_1$) with other relativistic images packed together
($\theta_\infty$) given by $s=\theta_\infty - \theta_1$:
\begin{center}
\begin{center} \bf {Table I:} Einstein ring and the relativistic images for JNW wormhole
\end{center}
\vspace {.25in}
\begin{tabular}{|c|c|c|c|c|} \hline
$\gamma$ &  $x_0$ , $\alpha_E$ (arcsec) , $\theta_E$ (arcsec) &
$\theta_1$ ($\mu$ arcsec) & $\theta _\infty$ ($\mu$ arcsec) &
  $s = \theta _\infty - \theta _1 $($\mu$ arcsec)\\ \hline
    1.0 &   178457 , 2.31166 , 1.15582  & 16.885295 & 17.792393 & 0.907098  \\
    1.5 &   218564 , 2.83119 , 1.41558  & 25.921124 & 26.712847 & 0.791723  \\
    2.0 &   252376 , 3.26919 , 1.63458  & 34.843479 & 35.663729 & 0.820249  \\
  \hline
\end{tabular}
\end{center}
\vspace {.25in} \noindent Finally we note that the Einstein angle
for the JNW metric is given by $\theta_{E}^{JNW} = \sqrt{\frac{4
\gamma m D_{ds}}{D_d D_s}}$ while that for the Ellis is
$\theta_{E}^{Ellis} = (\frac{\pi a^2D_{ds}}{4D_d^2D_s})^{(2/3)}$.
This readily shows that the ratio
$\frac{\theta_E^{JNW}}{\theta_E^{Sch}} \approx 1.41$ ($\gamma=2$)
and $\frac{\theta_E^{Ellis}}{\theta_E^{Sch}} \approx 10^{-17}$,
indicating that the Einstein ring of the JNW wormhole is much larger
than the Ellis wormhole. Furthermore, with the throat parameter
$a=10km$, the Einstein ring of the Ellis wormhole is found to be
$\theta_{E}^{Ellis}\approx 1.07 \times 10^{-11}$ micro-arcsecond and
therefore it is too small to be visible.

\par
To see how the qualitative features of the lens parameters of both
wormholes are similar, we compare the magnification curves of the
JNW wormhole $(\gamma=2)$ with that of the Ellis wormhole. Fig.3 and
4 depict the plots of the tangential magnification $(\mu_t)$ and
total magnification $(\mu)$ as the function of the image position
$\theta$ near the angular radius of the Einstein ring. For both
wormholes the singularities in these curves give the tangential
critical curves (TCC). Similarly Fig.5 and 6 show the variation of
the radial magnification $(\mu_r)$ where the absence of the
singularity indicates that we do not have any radial critical curve
(RCC).

\begin{center}   \textbf{V.Conclusion}  \end{center}
\par
The search for the wormholes by studying the mirror images formed
due to the gravitational lensing is one of the most intriguing way
to prove the existence of the exotic matter in nature. In this paper
we have explicitly calculated the deflection angle using the
methodology developed by Amore and Diaz for the JNW and Ellis
wormholes respectively. It is shown that the Einstein ring of the
JNW wormhole is much larger in comparison to that of the Ellis
wormhole. Furthermore, the JNW wormhole exhibits the relativistic
images in contrast to the Ellis wormhole which does not have any
photon sphere. A careful analysis of this pattern of the images may
shed some light on possible signature of the exotic lensing object
which gives rise to the wormhole geometry.
\begin{center}
{\bf Acknowledgement}
\end{center}

TKD thanks the University Grants Commission, New Delhi and SS thanks
Department of Science and Technology, New Delhi for partial support.
SS thankfully acknowledges the Visiting Associateship of S N Bose
National Centre for Basic Sciences, Kolkata.

\bibliographystyle{plain}

\begin{thebibliography}{5.8 in}
\bibitem[1] {} J N Hewitt \textit{et al}, Nature (London)
\textbf{333}, 537 (1988)
\bibitem[2] {} J Wambsganss, "Gravitational Lensing in Astronomy",
astro-ph/9812021
\bibitem[3] {} P Schneider, J Ehlers and E E Falco, \textit{Gravitational
Lenses} (Springer-verlag, Berlin, 1992)
\bibitem[4] {} K S Virbhadra and G F R Ellis, Phys Rev D \textbf {62},
084003 (2000)
\bibitem[5] {} K S Virbhadra and G F R Ellis, Phys Rev D \textbf {65},
103004 (2002)
\bibitem[6] {} K S Virbhadra, D Narasimha and S M Chitre, Astr \&
Astrophys \textbf {337}, 1 (1998)
\bibitem[7]  {} V Bozza, Phys Rev D \textbf {66}, 103001 (2002)
\bibitem[8] {} M S Morris and K S Thorne, Am J Phys \textbf {56}, 395 (1988)
\bibitem[9] {}  K K Nandi, Y Z Zhang, A V Zakharov, Phys Rev D \textbf {74}, 024020 (2006)
\bibitem[10] {} C Armend\'{a}riz-Picon, Phys Rev D \textbf {65}, 104010 (2002)
\bibitem[11] {} J G Cramer, R L Forward, M S Morris, M Visser, G Benford, G A Landis,
Phys Rev D \textbf {51}, 3117 (1995)
\bibitem[12] {} M Visser, \emph{Lorentzian wormholes}, AIP Press, New York 1995
\bibitem[13] {} S Weinberg, \emph{Gravitation and Cosmology: principles and
applications of the general theory of relativity}, John Wiely \&
Sons, New York 1972
\bibitem[14] {} P Amore and S A Diaz, Phys Rev D  \textbf {73}, 083004 (2006)
\bibitem[15]  {} S Frittelli, T P Kling and T Newman, Phys Rev D \textbf
{61},
064021 (2000)
\bibitem[16]  {} A I Janis, E T Newman and J Winnicour, Phys Rev Lett \textbf
{20}, 878 (1968)
\bibitem[17]  {} M Wynman, Phys Rev D \textbf
{24}, 839 (1981)
\bibitem[18]  {} H G Ellis, J Math Phys, \textbf {14}, 104 (1973)
\bibitem[19]  {} V Perlick, Phys Rev D \textbf {69}, 064017 (2004)
\bibitem[20] {} P Amore, S Arceo and F M Fern\'{a}ndez, Phys Rev D  \textbf {74},
083004 (2006)
\end{thebibliography}

\pagebreak

\begin{figure}
\begin{center}
\centerline{\epsffile{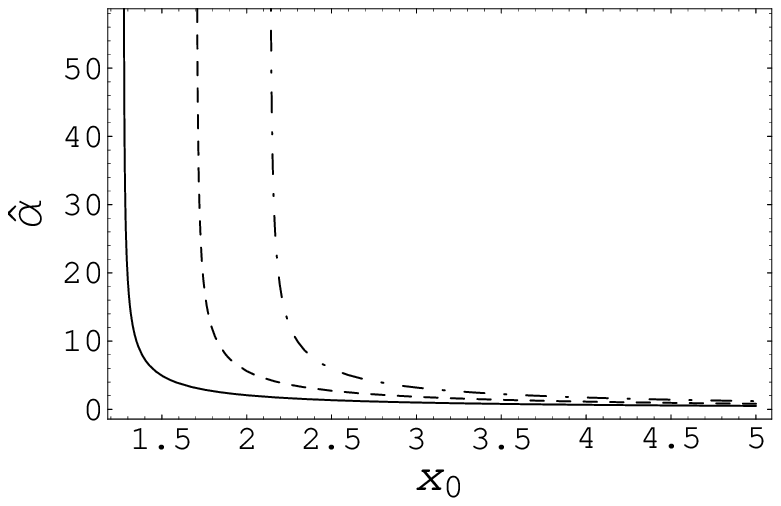}}
\end{center}
\noindent {\small {\bf Fig.1:} Variation of the deflection angle (in
degree) of the JNW metric with the distance of closest approach
$x_0$ ($x_0=\frac{r_0}{2m}$) for $\gamma=1$ (continuous curve),
$\gamma=1.5$ (dashed curve) and $\gamma=2$ (dot-dashed curve).}
\end{figure}

\begin{figure}
\begin{center}
\centerline{\epsffile{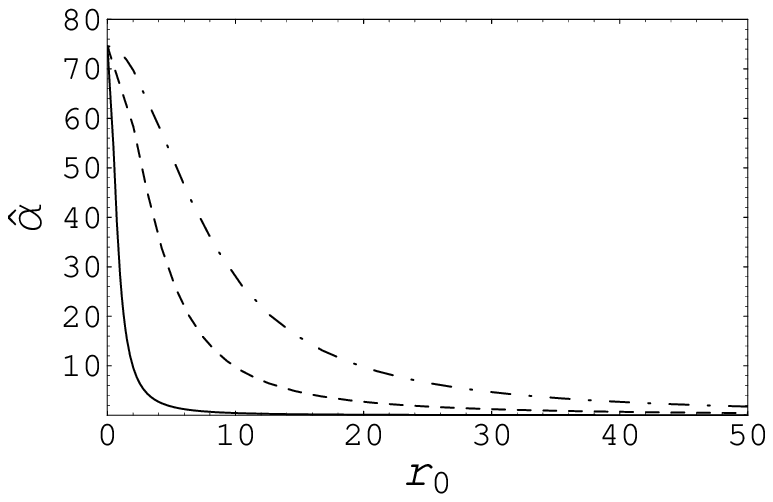}}
\end{center}
\noindent {\small {\bf Fig.2:} Variation of the deflection angle (in
degree) with the distance of closest approach $r_0$ (in km) for the
Ellis wormhole with throat parameter, $a=1$ km (continuous line),
$a=5$ km (dashed line), $a=10$ km (dot-dashed line)}.
\end{figure}

\begin{figure}
\begin{center}
\centerline{\epsffile{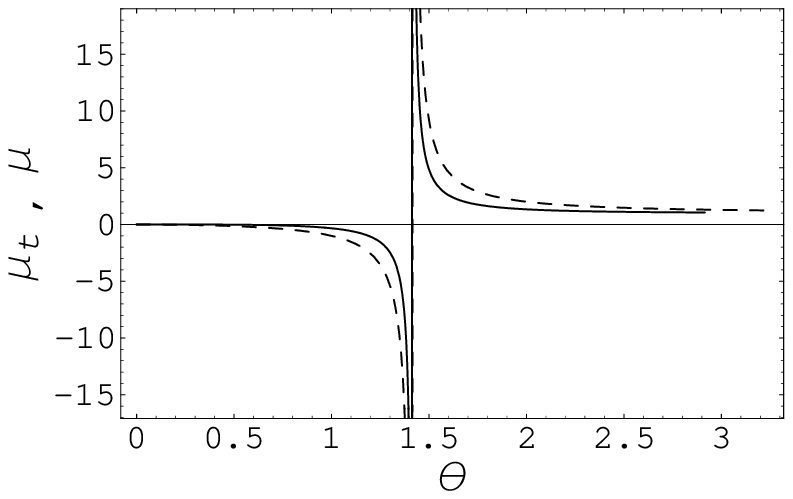}}
\end{center}
\noindent {\small {\bf Fig.3:} The tangential magnification $\mu_t$
(dashed curve) and total magnification $\mu$ (continuous curve) of
the JNW wormhole with $\gamma=2$ as the function of the image
position $\theta$  (in arcsecond) near the Einstein angle.}
\end{figure}

\begin{figure}
\begin{center}
\centerline{\epsffile{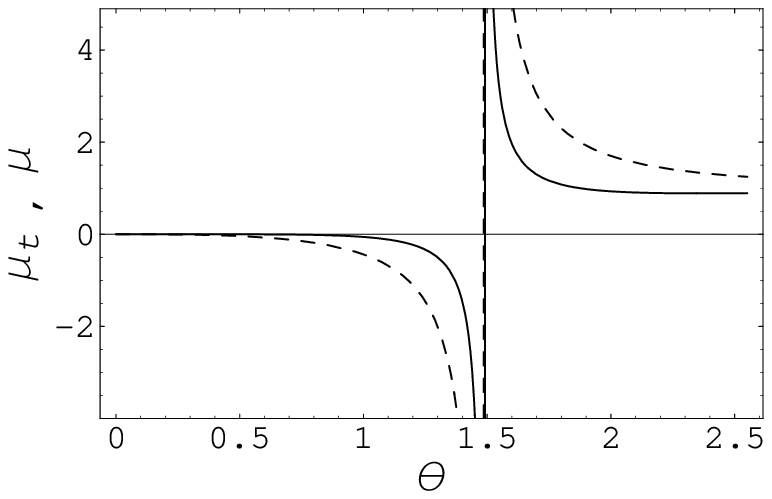}}
\end{center}
\noindent {\small {\bf Fig.4:} The tangential magnification $\mu_t$
(dashed curve) and total magnification $\mu$ (continuous curve) of
the Ellis wormhole ($a=10$ km) as the function of the image position
$\theta$ (in micro-arcsecond) near the Einstein angle.}
\end{figure}

\begin{figure}
\begin{center}
\centerline{\epsffile{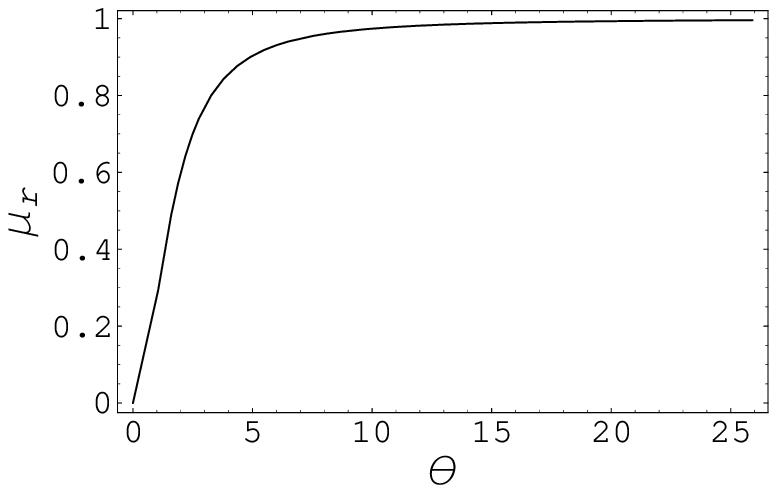}}
\end{center}
\noindent {\small {\bf Fig.5:} The radial magnification $\mu_r$
plotted against $\theta$ (in arcsecond) for the JNW wormhole with
$\gamma=2$.}
\end{figure}

\begin{figure}
\begin{center}
\centerline{\epsffile{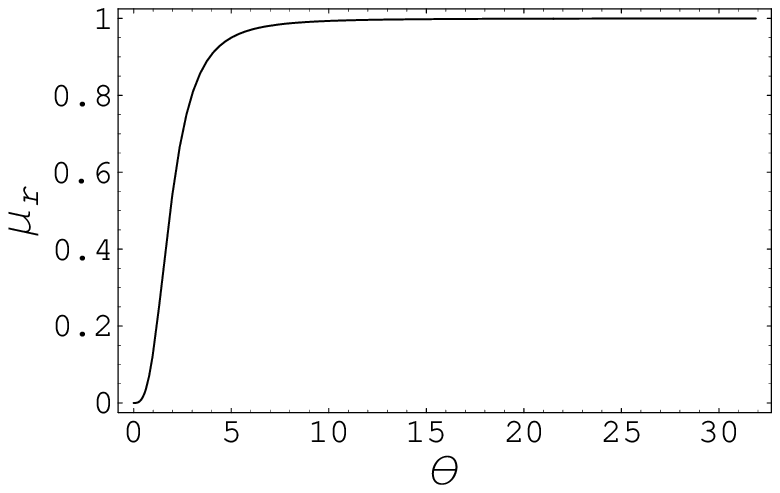}}
\end{center}
\noindent {\small {\bf Fig.6:} The radial magnification $\mu_r$
versus $\theta$ (in microarcsecond) for the Ellis wormhole ($a=10$
km).}
\end{figure}

\end{document}